\documentclass[a4paper,conference]{IEEEtran}
\IEEEoverridecommandlockouts
\usepackage{cite}
\usepackage{amsmath,amssymb,amsfonts}
\usepackage{algorithmic}
\usepackage{graphicx}
\usepackage{textcomp}
\usepackage{xcolor}
\usepackage{upgreek}
\def\BibTeX{{\rm B\kern-.05em{\sc i\kern-.025em b}\kern-.08em
    T\kern-.1667em\lower.7ex\hbox{E}\kern-.125emX}}
\usepackage[T1]{fontenc}
\usepackage{cite}
\usepackage{hyperref}
\usepackage{booktabs}
\usepackage{siunitx}
\usepackage{bm}
\def\BibTeX{{\rm B\kern-.05em{\sc i\kern-.025em b}\kern-.08em
    T\kern-.1667em\lower.7ex\hbox{E}\kern-.125emX}}
\newcommand{\Ethz}{E_{\mathrm{THz}}}
\newcommand{\Ebias}{E_{\mathrm{bias}}}

\newcommand{\dif}{\mathrm{d}}
\begin{document}

\title{Self-Supervised Noise2Noise-Enhanced Denoising for Continuous-Scan Air-Plasma THz Spectroscopy\\

\thanks{This work was supported by the German Research
Foundation (DFG) under Project–ID 287022738 TRR 196 for Project S02 and Project S03.}
}

\author{
    \IEEEauthorblockN{Adam Umra$^\star$, Oways Alsoloh$^\star$, Oliver Nagy$^\dagger$, Aydin Sezgin$^\star$, Clara Saraceno$^\dagger$}
    \IEEEauthorblockA{$^\star$Institute of Digital Communication Systems, Ruhr University Bochum, Germany\\
    $^\dagger$Photonics and Ultrafast Laser Science, Ruhr University Bochum, Germany\\
    \{adam.umra, oways.alsoloh, oliver.nagy, aydin.sezgin, clara.saraceno\}@rub.de}
}

\maketitle

\begin{abstract}
Terahertz time-domain spectroscopy (THz-TDS) based on air-plasma
generation and balanced air-biased coherent detection offers gap-free
broadband coverage, but individual continuous-scan traces are strongly
affected by pulse-to-pulse fluctuations and electronic noise. Reaching
a useful signal-to-noise ratio therefore requires averaging multiple
traces, which directly increases measurement time. We propose a learned
denoising approach that recovers high-quality THz waveforms from as few
as one complete continuous delay sweep, referred to here as a
single-scan trace. A compact one-dimensional residual U-Net is trained
using two complementary strategies: a reference-supervised baseline
that maps individual noisy traces to long-average reference waveforms,
and a Noise2Noise approach that learns from pairs of independently
acquired noisy traces without requiring a clean training target.
Averaging the predictions of both models reduces systematic bias and
yields a trace-reduction factor of approximately $5.4\times$ at $K=1$,
meaning that one denoised trace achieves the reconstruction accuracy of
averaging approximately five raw traces. The Noise2Noise model alone
achieves $4.9\times$, outperforming both the reference-supervised
baseline ($4.6\times$) and classical Wiener filtering ($3.2\times$).
These results show that self-supervised learning from repeated noisy
measurements can support faster continuous-scan THz-TDS without
hardware modification.
\end{abstract}


\section{Introduction}
\label{sec:intro}

The terahertz (THz) spectral range, located between microwave and
infrared frequencies, is attracting growing interest for emerging
high-frequency wireless communication as well as for spectroscopy,
sensing, imaging, and non-destructive testing~\cite{thz_communications, thz_sensing, karacora2025THz}.
Many materials exhibit characteristic rotational, vibrational, or
collective excitations in this frequency range, making THz radiation a
sensitive and non-ionizing probe of material properties. Terahertz
time-domain spectroscopy (THz-TDS)~\cite{koch2023THZTDS} is particularly
well suited for such measurements because it directly records the
electric field of a broadband THz pulse, providing both amplitude and
phase information after Fourier transformation.

Air-plasma-based THz-TDS systems are attractive because generation and
detection occur in ambient gas rather than in crystalline media,
enabling broad, gap-free spectral coverage that is not limited by
phonon resonances of electro-optic crystals~\cite{airplasma_review}.
Detection schemes such as air-biased coherent detection provide
broadband sensitivity and high dynamic range for measuring ultrashort
THz transients~\cite{abcd_ref}. A practical limitation, however, is
acquisition speed: conventional THz-TDS measurements often require
averaging many repeated laser shots or scans to suppress shot-to-shot
laser fluctuations, electronic noise, timing jitter, and baseline
variations. This increases measurement time and becomes restrictive for
applications requiring rapid feedback, repeated measurements,
high-throughput screening, or measurements over many sample positions.

The central problem addressed in this work is the recovery of
high-quality THz time-domain traces from a small number of noisy
single-scan traces or low-average measurements. Classical denoising methods,
such as post-acquisition averaging and Wiener
filtering~\cite{wiener1949Filter}, can improve noisy traces but rely on
additional measurements or fixed assumptions about the signal and noise
statistics. Neural-network-based denoising offers a data-driven
alternative, but standard supervised training typically requires paired
noisy inputs and clean, or at least heavily averaged, reference targets,
which partly reintroduces the acquisition burden that denoising is
intended to reduce.

This work exploits a feature naturally available in continuous-scan
THz-TDS: repeated scans of an unchanged sample provide multiple noisy
realizations of the same underlying waveform. This setting enables the
Noise2Noise approach~\cite{lehtinen2018N2N}, in which a model is trained using pairs of
noisy observations instead of clean targets. Provided that the paired
traces represent the same physical state and that their noise is
approximately independent and zero-mean, minimizing the expected
training loss recovers the underlying waveform. Noise2Noise therefore
provides a practical route to THz waveform denoising without requiring
a dedicated long-average target for training.

The main contribution of this work is the adaptation and evaluation of
this principle for shot-efficient continuous-scan air-plasma THz-TDS.
We train a compact one-dimensional residual U-Net using two
complementary strategies: a reference-supervised model that maps an
individual noisy trace to a long-average reference, and a Noise2Noise
model trained only on pairs of noisy traces acquired from the same
sample state. Comparing the two strategies shows whether useful
denoising can be learned directly from repeated noisy measurements,
whereas their ensemble combines the complementary estimates. The study
thus evaluates Noise2Noise under the strong shot-to-shot fluctuations
encountered in air-plasma THz-TDS, while avoiding clean training targets
for the self-supervised model.

The proposed denoisers are evaluated against raw averaging and a
classical Wiener-filter baseline on a held-out acquisition. Performance
is quantified using root-mean-square error, signal-to-noise ratio, and
an effective trace-reduction factor that links reconstruction quality
directly to acquisition-time reduction. The results show that learned
denoising, particularly the Noise2Noise-based approach, improves
few-trace THz waveform reconstruction relative to raw averaging and
Wiener filtering, supporting faster THz-TDS acquisition without
additional hardware.
\section{THz Generation and Detection}
\label{sec:thz_system}

The measurements are performed with a continuous-scan air-plasma THz-TDS system using two-color THz generation and balanced air-biased coherent detection (ABCD). For generation, a fundamental optical pulse at angular frequency $\omega$ (\SI{800}{\nano\meter}) and its second harmonic at $2\omega$ (\SI{400}{\nano\meter}) are co-focused in ambient air~\cite{kim_2color, balanced_abcd}. The resulting plasma filament produces a temporally asymmetric driving field whose transient photocurrent radiates a broadband THz pulse,
\begin{equation}
  \Ethz(t) \propto \frac{\dif J(t)}{\dif t}.
  \label{eq:photocurrent}
\end{equation}
Since the emission process is not limited by crystal phase matching, air-plasma sources provide continuous broadband spectra extending from the sub-\si{\THz} range to tens of \si{\THz}.

On the detection side, ABCD reads out the THz field through a third-order nonlinear interaction between the THz transient, an optical probe pulse, and a static bias field~\cite{abcd_dai}. The detected second-harmonic intensity contains a heterodyne cross-term proportional to the THz field,
\begin{equation}
  I_{\mathrm{het}}(t) \propto \Ebias \Ethz(t),
  \label{eq:abcd_signal}
\end{equation}
which preserves the signed time-domain electric-field waveform, from which spectral amplitude and phase are obtained. In the balanced ABCD configuration used here, the two bias-polarity components are encoded simultaneously in orthogonal polarization channels and measured by separate avalanche photodiodes (APDs)~\cite{ohrt2024THz}. This provides shot-to-shot common-mode rejection without requiring bias modulation over consecutive laser shots, which is essential for single-scan operation. The pump--probe delay is swept using a continuously oscillating delay stage rather than a step-and-integrate scan. The boxcar-integrated APD signals are digitized on the fly at a \SI{1}{\kilo\hertz} laser repetition rate, so each forward or backward sweep yields one complete THz waveform in approximately \SI{0.5}{\second}~\cite{ohrt2024THz}. This greatly improves acquisition speed, but each trace contains only minimal averaging and is therefore strongly affected by trace-to-trace noise. This trade-off motivates the denoising approach introduced in Section~\ref{sec:method}.
\section{Dataset}
\label{sec:data}

\subsection{Acquisitions}
\label{ssec:acquisitions}

All data were recorded with the balanced-ABCD continuous-scan system described in Section~\ref{sec:thz_system}.  Three independent acquisitions are available, each stored as a matrix of single-scan trace waveforms sampled at a stage step of \SI{1.3}{\micro\meter}, which corresponds to a round-trip time increment of $\Delta t = 2 \times \SI{1.3}{\micro\meter} / c \approx \SI{8.67}{\femto\second}$. The first two acquisitions contain $N = 200$ traces each; the third contains $N = 100$ traces, yielding 500 single-scan trace waveforms in total.

\subsection{Preprocessing}
\label{ssec:preprocessing}

Each raw waveform matrix is preprocessed in three steps. First, a per-trace DC offset is removed by subtracting the trace mean, which eliminates slow electronic drift between shots. Second, every odd-indexed trace is time-reversed to account for the bidirectional stage motion: odd traces are recorded on the return sweep and must be reflected before they can be compared with even traces. Third, all traces are multiplied by $-1$ to align the signal polarity with convention.

After preprocessing, a window of $L = 80$ samples ($\approx \SI{693}{\femto\second}$) is extracted from each trace, centered on the sample of maximum absolute value in the corresponding long-average waveform.  The long-average $\bar{x}$ of an acquisition is the mean over all $N$ preprocessed traces and serves as the reference ground truth for evaluation.  All amplitudes are globally normalized by the standard deviation of the training waveforms.

\subsection{Training and Validation Split}
\label{ssec:split}

The two 200-trace acquisitions are used for training and the 100-trace acquisition is held out as the validation set.  The long-average of the validation acquisition is never used during training; it serves exclusively as the reference for computing RMSE and SNR.

\section{Network Architecture and Training}
\begin{figure*}
    \centering
    \includegraphics[width=0.8\linewidth]{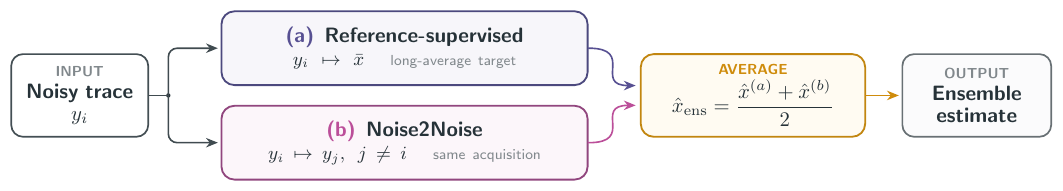}
    \caption{Proposed denoising framework. Two independently trained residual U-Nets use reference-supervised and Noise2Noise targets, respectively. Their predictions are averaged to obtain the ensemble estimate.}
    \label{fig:denoising_framework}
\end{figure*}

Figure~\ref{fig:denoising_framework} gives an overview of the complete
denoising framework. Both branches process the same noisy input using
independently trained instances of the same one-dimensional residual
U-Net. Variant~(a) is trained against the long-average reference,
whereas variant~(b) uses a different noisy trace from the same
acquisition as its target. During inference, the predictions of both
models are averaged to obtain the final estimate.
\label{sec:method}
\subsection{One-Dimensional Residual U-Net}
\label{ssec:architecture}

Both learned denoisers share a common one-dimensional residual
U-Net~\cite{unet_original, unet_1d} backbone.  The network takes a
single noisy trace of length $L = 80$ as input and outputs a denoised
trace of the same length.

The encoder comprises three successive convolutional blocks, each
followed by average pooling with stride 2, reducing the temporal
dimension from 80 to 10 while expanding the channel depth:
$1 \!\to\! 32 \!\to\! 64 \!\to\! 128$.
A bottleneck block operates at the coarsest resolution
($128$ channels, length $10$).
The decoder mirrors this structure with nearest-neighbour upsampling;
skip connections concatenate the corresponding encoder feature maps
before each decoder block, restoring the channel sequence
$128 \!\to\! 64 \!\to\! 32$, followed by a $1{\times}1$ convolution
to produce the single-channel output.
Each convolutional block consists of two \texttt{Conv1d} layers with
kernel size 5 and same-padding, each followed by Group
Normalization~\cite{groupnorm} and a GELU activation~\cite{gelu}.
The network output is added back to the input as a global residual
connection, so the network learns only the correction rather than the
full waveform.  The total parameter count is approximately 467\,000.

\subsection{Training Variants and Objectives}
\label{ssec:variants}

\textbf{(a) Reference-supervised.}
Each noisy training trace $y_i$ is paired with the long-average
$\bar{x}$ of its acquisition file, which serves as the supervised
target.  The model is trained to minimize the combined
loss~\eqref{eq:totalloss}.  This is the standard supervised
baseline in the present setting; its principal limitation is that
only two distinct target shapes are available (one per training
file), so the model must generalize from a small set of reference
waveforms. 

For variant~(a), two loss terms are combined.  The primary term is
the time-domain mean-squared error (MSE) between the prediction
$\hat{x}$ and the target $x$:
\begin{equation}
  \mathcal{L}_{\mathrm{time}} = \frac{1}{L}\|\hat{x} - x\|^2.
  \label{eq:ltime}
\end{equation}
An auxiliary spectral loss penalizes log-magnitude errors.  The
one-sided DFT magnitude is normalized by the target spectral peak and
expressed in decibels, with a floor at $-40$\,dB to suppress
numerical noise in spectral nulls.  The combined loss is
\begin{equation}
  \mathcal{L} = \mathcal{L}_{\mathrm{time}} + \lambda\,\mathcal{L}_{\mathrm{spec}},
  \quad \lambda = 0.1.
  \label{eq:totalloss}
\end{equation}

\textbf{(b) Noise2Noise.}
Following~\cite{lehtinen2018N2N}, two independently drawn traces
$(y_i, y_j)$ from the same acquisition file are used as
input--target pairs, with $i \neq j$ drawn uniformly at random.  No
long-average reference is required at training time.  Because the
noise contributions are independent and approximately zero-mean, the
expected value of $y_j$ given the underlying clean signal equals
$\bar{x}$, so MSE minimization converges to the same fixed point as
fully supervised training.  This variant is therefore self-supervised:
it exploits the repeated measurements already present in the
continuous-scan acquisition without requiring any additional
reference.

For variant~(b), only $\mathcal{L}_{\mathrm{time}}$ is used
($\lambda = 0$), as the Noise2Noise target is itself noisy and
spectral shaping would amplify the target noise.

\subsection{Ensemble}
\label{ssec:ensemble}

The two trained models exhibit complementary error characteristics:
variant~(a) is anchored to the long-average shapes seen during
training, while variant~(b) is guided solely by the pairwise noise
structure.  An ensemble prediction is formed by averaging their
outputs,
\begin{equation}
  \hat{x}_{\mathrm{ens}} = \tfrac{1}{2}
    \bigl(\hat{x}^{(a)} + \hat{x}^{(b)}\bigr),
  \label{eq:ensemble}
\end{equation}
which reduces the variance contribution from each model's individual
bias.

\subsection{Optimization}
\label{ssec:optim}

Both variants are trained with the Adam optimizer at an initial
learning rate of $2 \times 10^{-3}$, decayed to zero with a cosine
schedule over the full training run.  The batch size is 32.
Variant~(a) trains for 400 epochs over the full 400-trace training
set ($1.6 \times 10^5$ training samples total); variant~(b) draws
4\,000 randomly sampled intra-file pairs per epoch for 40 epochs,
matching the same total step count.

\section{Numerical Results}
\label{sec:results}

\subsection{Evaluation Protocol}
\label{ssec:eval_protocol}

The held-out 100-trace acquisition serves as the test set; its
long-average $\bar{x}$ is the reconstruction target and is never
observed during training.  For a denoiser producing per-trace
predictions $\{\hat{x}_i\}$, the $K$-trace estimate is the mean over
a randomly drawn subset of size $K$.  Quality is measured by the
root-mean-square error (RMSE) to $\bar{x}$, averaged over 500 random
subsets drawn without replacement.  The trace-reduction factor
\begin{equation}
  \rho(K) = K_{\mathrm{eq}}(K) \,/\, K
  \label{eq:rho}
\end{equation}
quantifies how many raw traces $K_{\mathrm{eq}}$ would be needed to
reach the same RMSE as $K$ denoised traces, estimated by log-log
interpolation of the raw $K$-sweep curve.

\subsection{Qualitative Comparison}
\label{ssec:qualitative}

\begin{figure}[t]
  \centering
  \includegraphics[width=0.8\columnwidth]{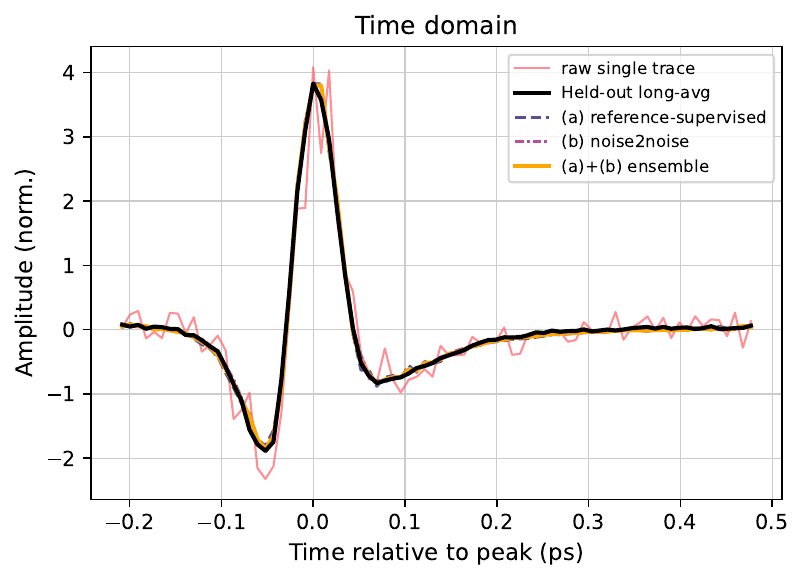}\\[4pt]
  \includegraphics[width=0.8\columnwidth]{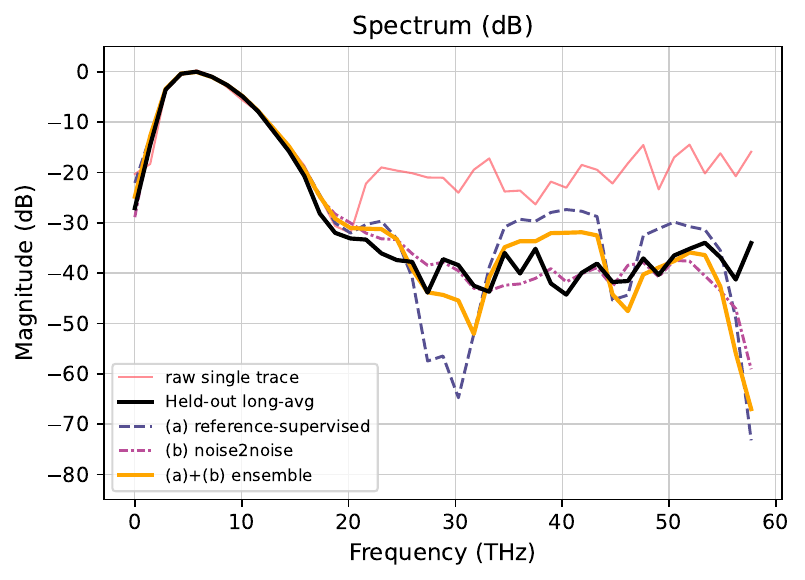}
  \caption{Qualitative comparison on the held-out acquisition.
\textit{Top:} Time-domain waveforms for the long-average reference,
a representative raw single trace, and the mean predictions of the
reference-supervised, Noise2Noise, and ensemble models.
\textit{Bottom:} Corresponding one-sided magnitude spectra, normalized
to the peak of the long-average reference.}
  \label{fig:overlay_td_spec}
\end{figure}

Figure~\ref{fig:overlay_td_spec} shows the mean denoised waveform of
each method alongside the held-out reference and a representative raw
single trace.  In the time domain the raw trace is dominated by
trace-to-trace fluctuations that largely obscure the THz pulse shape,
while both learned denoisers and their ensemble closely reproduce the
reference waveform.  The spectral panel makes the benefit quantitative:
the raw single trace raises the noise floor by approximately
\SI{20}{\deci\bel} across the full bandwidth, whereas all denoised
curves track the reference spectrum within a few decibels up to the
system's spectral limit.  Variant~(b) achieves a lower RMSE than
variant~(a), demonstrating that the self-supervised Noise2Noise
approach matches or exceeds the reference-supervised baseline
without ever observing a clean target.  The ensemble further reduces the error by
averaging the complementary bias components of the two models.

\begin{figure}[t]
  \centering
  \includegraphics[width=0.8\columnwidth]{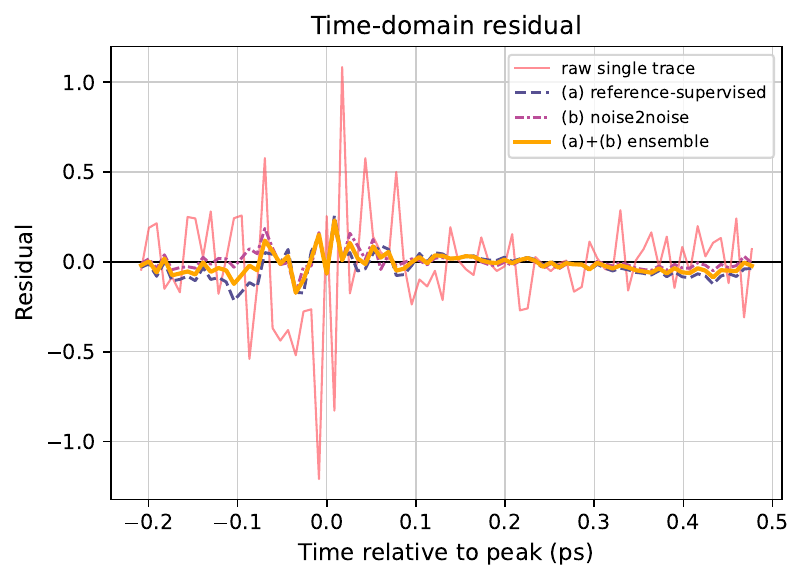}\\[4pt]
  \includegraphics[width=0.8\columnwidth]{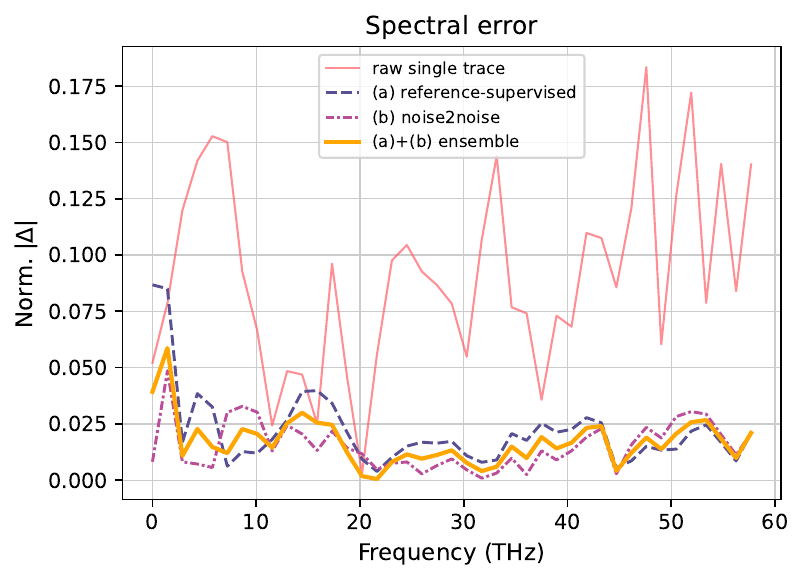}
  \caption{Residual analysis on the held-out acquisition.
    \emph{Top}: time-domain residual (denoised mean $-$ reference).
    \emph{Bottom}: normalized spectral error
    $|\mathcal{F}\{\hat{x}\} - \mathcal{F}\{\bar{x}\}|\,/\,\max|\mathcal{F}\{\bar{x}\}|$.}
  \label{fig:overlay_res}
\end{figure}

Figure~\ref{fig:overlay_res} confirms these observations in the residual
domain.  The time-domain residual of the raw single trace is an order
of magnitude larger than that of any denoised estimate, and the
spectral error panel shows that the residual of the raw trace is
spectrally broadband, whereas the denoised residuals are
concentrated at low frequencies where model bias is largest.  The
ensemble achieves the smallest residual in both domains.

\subsection{Few-trace Performance}
\label{ssec:ksweep}

\begin{figure}[t]
  \centering
  \includegraphics[width=0.8\columnwidth]{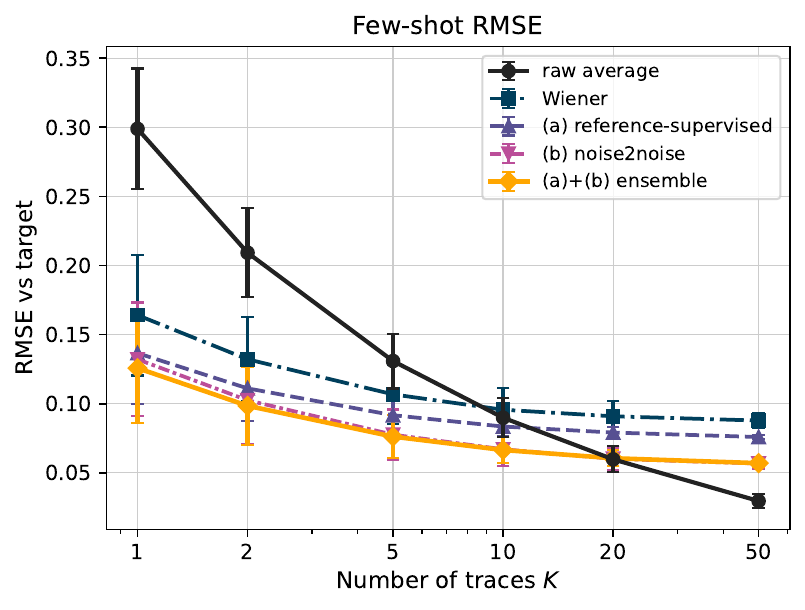}\\[4pt]
  \includegraphics[width=0.8\columnwidth]{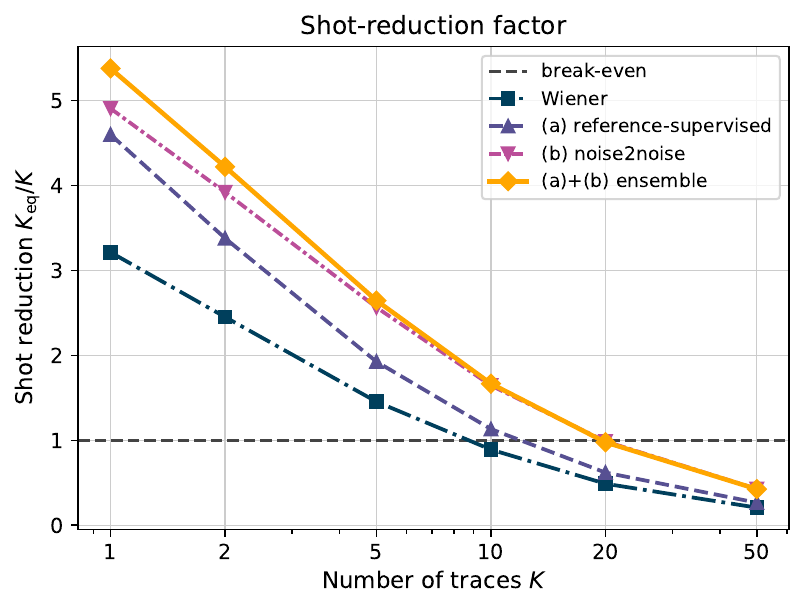}
  \caption{$K$-sweep evaluation on the held-out acquisition.
    \emph{Top}: RMSE vs.\ number of averaged traces~$K$; error bars
    show $\pm1$ standard deviation over 500 random subsets.
    \emph{Bottom}: trace-reduction factor $\rho(K)=K_{\mathrm{eq}}/K$;
    dashed line marks break-even with raw averaging.}
  \label{fig:ksweep}
\end{figure}

Figure~\ref{fig:ksweep} reports the RMSE and trace-reduction factor as
a function of~$K$.  In the RMSE panel three regimes are visible: for
$K \lesssim 4$ all learned denoisers sit well below the raw-average
curve; around $K \approx 5$ the raw average closes the gap as
$1/\sqrt{K}$ averaging reduces noise faster than the models' residual
bias floor; beyond $K \approx 10$ raw averaging is competitive with
or superior to the denoised estimates, which plateau at a fixed RMSE
floor.  The Wiener filter follows the same qualitative pattern but at
a higher bias floor than the neural methods.

The trace-reduction panel is the headline result.  At $K=1$, the
(a)+(b) ensemble achieves $\rho \approx 5.4$, meaning a single
denoised trace is equivalent in RMSE to averaging roughly five raw
traces.  Variant~(b) alone reaches $\rho \approx 4.9$ and variant~(a)
$\rho \approx 4.6$, with the Noise2Noise model outperforming the
reference-supervised baseline despite its lack of clean training targets.  The
classical Wiener filter achieves $\rho \approx 3.2$, well below the
neural denoisers but still a clear improvement over raw averaging.  The advantage is most pronounced at low~$K$ and
decays toward unity as $K$ grows, precisely the few-trace regime
where acquisition time is the limiting factor. In the acquisition configuration used here, one complete THz scan requires approximately $0.5$~s. Thus, achieving the reconstruction quality of 5.4 raw traces by conventional averaging requires approximately $2.7$~s, whereas the proposed ensemble reaches comparable RMSE from a single $0.5$~s scan.

Table~\ref{tab:headline} consolidates the per-method reconstruction
quality at $K{=}1$.  The SNR gain column reports the noise-power
improvement of each denoised single trace relative to a raw single
trace, $\text{SNR\,gain} = 20\log_{10}(\mathrm{RMSE}_{\mathrm{raw}}/
\mathrm{RMSE}_{\mathrm{method}})$.

\begin{table}[t]
  \centering
  \caption{single-scan trace($K{=}1$) reconstruction quality on the
    held-out acquisition.  Lower RMSE and higher SNR gain are
    better; the raw single trace defines the $0$\,dB reference.}
  \label{tab:headline}
  \small
  \setlength{\tabcolsep}{6pt}
  \begin{tabular}{lcc}
    \toprule
    Method                  & RMSE      & SNR gain (dB) \\
    \midrule
    Raw single trace         & 0.293     & 0.00          \\
    Wiener filter           & 0.164     & 5.07          \\
    (a) Reference-supervised& 0.137     & 6.63          \\
    (b) Noise2Noise         & 0.132     & 6.91          \\
    (a)+(b) Ensemble        & 0.126     & 7.31          \\
    \bottomrule
  \end{tabular}
\end{table}

\subsection{Generalization and Limitations}
The evaluation uses an acquisition-level holdout: the 100-trace test acquisition is not observed during training. It therefore measures generalization to an unseen acquisition, but not to a fundamentally different sample, waveform, or experimental configuration. Since the training set contains only two acquisitions, the range of waveform shapes and measurement conditions represented during training remains limited. The Noise2Noise formulation assumes that paired traces contain the same underlying waveform and approximately independent, zero-mean noise~\cite{lehtinen2018N2N}. Correlated baseline drift, systematic timing errors, or changes in the sample between paired scans violate these assumptions and may be learned as part of the signal. For time-varying measurements, the model may suppress genuine temporal changes if they resemble trace-to-trace fluctuations. Performance may also degrade for waveform shapes, amplitudes, spectral features, or noise levels that differ substantially from those represented during training. Broader validation across samples and acquisition conditions is therefore required before deployment as a general-purpose THz denoiser.

\section{Conclusion}
\label{sec:conclusion}

We have shown that learned single-scan trace denoising can substantially
reduce the number of continuously scanned air-plasma THz traces required
for high-quality reconstruction.  An ensemble of a supervised residual
U-Net and a Noise2Noise-trained partner achieves a trace-reduction factor
of approximately $5.4\times$ at $K{=}1$, compared with $3.2\times$ for
Wiener filtering, with the largest gains in the few-trace regime
($K \lesssim 4$). The Noise2Noise model performs competitively without clean targets,
showing that repeated noisy acquisitions of the same physical state are
sufficient for training.  This avoids dedicated long-average reference
measurements, and the ensemble further reduces systematic bias at no
additional data cost.

\bibliographystyle{IEEEtran}
\bibliography{bibliography}

\end{document}